 \documentstyle[preprint,prl,aps]{revtex}
\begin{document}
\draft
\title{Self-similarity limits of genomic signatures}
\author{Zuo-Bing Wu} 
\address{ State Key Laboratory of Nonlinear Mechanics, 
Institute of Mechanics, Academia Sinica, 
Beijing 100080, China}
\maketitle

\begin{abstract}
It is shown that metric representation of DNA sequences is
one-to-one. By using the metric representation method,
suppression of nucleotide strings in the DNA sequences is determined. 
For a DNA sequence,
an optimal string length to display genomic signature in
chaos game representation is obtained by eliminating effects of
the finite sequence. The optimal string length is further
shown as a self-similarity limit in computing information dimension.
By using the method, self-similarity limits of bacteria complete
genomic signatures are further determined.
\end{abstract}

\section{Introduction}
Along with an increasing amount of  DNA sequences extracted from experiments,
it is important to develop methods for extracting meaningful information from
the one-dimensional symbolic sequences composed of the four letters `A', `C', 
`G' and `T' (or `U'). 
To detect similarity in DNA sequences, scatter plots\cite{GM} are introduced
to make classification of cytochromes and illustrate a dendrogram.
From a comparison of a pair of duplicated genes by a distance matrix, 
evolutionary relationship of three primary kingdoms of life is inferred\cite{IKHOM}.
Due to investigating relative abundances of short oligonucleotides in subsequences,
genomic signature phenomenon and derivation of partial-ordering relationships 
among bacterial genomes are proposed\cite{KMC}. The genomic signature describes that
the difference of dinucleotide relative abundance values within a single
genome is larger than that between distinct genomes.
Chaos game representation (CGR)\cite{Jeffrey}, which generates 
a two-dimensional square from a one-dimensional sequence, provides a technique to 
visualize the composition of DNA sequences. 
By composing the CGR and short-sequence representation methods, 
the evolution of species-type specificity in mitochondral genomes
is analyzed\cite{HS}. In terms of the CGR method, it is shown that
the main characteristics of the whole genome can be exhibited by
 its subsequences\cite{DGVFF}. 
 The genomic signature is extended to describe characteristics of CGR images. 
By making a Euclidean metric between two CGR images,
classification of species in three primary kingdoms is discussed. 

Recently, metric representation (MR)\cite{Wu}, which is borrowed from
the symbolic dynamics, makes an ordering of subsequences in a plane.
The MR method is an extension of CGR.
Suppression of certain nucleotide strings in the DNA sequences leads to
a self-similarity of pattern seen in the MR of DNA sequences.
In this paper, first, we show that the MR is one-to-one. Due to the MR method,
we determine suppression of nucleotide strings in DNA sequences. Then,
eliminating effects of finite sequences on suppression of nucleotide strings,
 we give an optimal string length to display genomic signature. 
  Moreover, we plot information function versus string lengths to 
  determine self-similarity limits in MR images.
  Using the method, we present self-similarity limits
  of bacteria complete genomic signatures.
 
\section{Suppression of nucleotide strings} 
 
For a given DNA sequence, we have
a one-dimensional symbolic sequence $s_1 s_2 \cdots s_i \cdots s_N$ ($s_i \in
\{A,C,G,T\}$).
In a two-dimensional MR, we take the correspondence of symbol $s_i$ to 
number $\mu_i$ or $\nu_i \in \{0,1\}$ and calculate the values 
($\alpha$, $\beta$) of all subsequences $\Sigma_m=s_1 s_2 \cdots s_m$ ($1 \le m \le N$).
The number $\alpha$ represented in base 3, between 0 and 1, is defined as 
 \begin{equation}
 \alpha  = 2\sum_{j=1}^m \mu_{m-j+1} 3^{-j} +3^{-m}= 2\sum_{i=1}^m \mu_i 3^{-(m-i+1)} +3^{-m},
 \label{eq3} 
 \end{equation} 
 where $\mu_i$ is 0 if $s_i \in \{A,C\}$ or 1 if $s_i \in \{G,T\}$. 
Similarly, the number $\beta$ is defined as
 \begin{equation}
 \beta = 2\sum_{j=1}^m \nu_{m-j+1} 3^{-j} +3^{-m}= 2\sum_{i=1}^m \nu_i 3^{-(m-i+1)} +3^{-m},
 \label{eq4}
 \end{equation}
 where $\nu_i$ is 0 if $s_i \in \{A,T\}$ or 1 if $s_i \in \{C,G\}$.  
According to (1) and (2), the one-dimensional symbolic sequence $s_1 s_2 \cdots s_N$
is partitioned into 4 kinds of subsequences, which correspond to points 
in 4 fundamental zones $A,C,G$ and $T$ of Fig.~1. Under left or right shift
operators, each zone can be further shrunk to less zones with a factor of $1/3^2$.
For an infinite sequence, this procedure can be defined as a fractal\cite{Mand}, 
which has a self-similarity. 
The subsequences with the same ending $k$-nucleotide string are labeled by $\Sigma_{(k)}$.
All subsequences $\Sigma_{(k)}$ correspond to points in the 
zone encoded by the $k$-nucleotide string.

{\bf Lemma 1} $(\alpha, \beta) \{ S(\Sigma_m) \} = 2( \mu_{m+1}, \nu_{m+1})/3
+ (\alpha, \beta) \{ \Sigma_m \}/3 $. $S$ is a left shift operator.

PROOF: Note that for the left shift operator, $S(\Sigma_m) = \Sigma_m s_{m+1}$.
From the definition (1) and (2), we can immediately obtain the result. 

{\bf Lemma 2} $(\alpha, \beta) \{ \Sigma_m \} = (\alpha, \beta) \{ G^{\infty} \Sigma_m \}$.

PROOF: When $m=1$, $\Sigma_1=s_1$ and $G^{\infty} \Sigma_1=S(G^{\infty})$.
By the Lemma 1, we can obtain $(\alpha, \beta) \{ G^{\infty} \Sigma_1 \}
=2( \mu_1, \nu_1)/3 + (\alpha, \beta) \{ G^{\infty} \}/3 =2( \mu_1, \nu_1)/3 + (1, 1)/3
= (\alpha, \beta) \{ \Sigma_1 \}$.
Suppose when $m=i$, we have 
$(\alpha, \beta) \{ \Sigma_i \} = (\alpha, \beta) \{ G^{\infty} \Sigma_i \}$.
For $m=i+1$, we have $\Sigma_{i+1}=\Sigma_i s_{i+1} =S(\Sigma_i)$ and
$G^{\infty} \Sigma_{i+1} =S(G^{\infty} \Sigma_i)$.
By the Lemma 1, we obtain $(\alpha, \beta) \{ \Sigma_{i+1} \}
=2( \mu_{m+1}, \nu_{m+1})/3 +(\alpha, \beta) \{ \Sigma_i \}/3 $ and
 $(\alpha, \beta) \{ G^{\infty}\Sigma_{i+1} \}
=2( \mu_{m+1}, \nu_{m+1})/3 +(\alpha, \beta) \{ G^{\infty}\Sigma_i \}/3 $.
So, using the supposition $(\alpha, \beta) \{ \Sigma_i \} = (\alpha, \beta) \{ G^{\infty} \Sigma_i \}$,
we can lead to $(\alpha, \beta) \{ \Sigma_{i+1} \} =
(\alpha, \beta) \{ G^{\infty}\Sigma_{i+1} \}$.

By the Lemma 2, each finite subsequence $\Sigma_m$ has 
a correspondent infinite sequence $G^{\infty} \Sigma_m$. Here, we define a set of the
infinite sequences as $\Sigma$.

{\bf Theorem 1} $(\alpha, \beta) : \Sigma \rightarrow \Lambda$
is one-to-one. $\Lambda$ is a set of points in the ($\alpha, \beta$) plane.

This means that given $\Sigma^1$, $\Sigma^2 \in \Sigma$,
if $\Sigma^1 \neq \Sigma^2$, then $(\alpha, \beta) \{\Sigma^1\}
\neq (\alpha, \beta) \{\Sigma^2\}$. We give a proof by contradiction.
Suppose $(\alpha, \beta) \{\Sigma^1\} = (\alpha, \beta) \{\Sigma^2\}$
and is marked as $P$ in the the ($\alpha, \beta$) plan.
For the zone including the point $P$, we encode it as two subsequences
$\Sigma_1^1$ and $\Sigma_1^2$ with the same mononucleotide. Then,
enlarge the zone  by a area factor of $3^2$, we can  obtain two
encoding subsequences $\Sigma_2^1$ and $\Sigma_2^2$ with the same dinucleotide.
Each enlarging process provides a right shift to two subsequences.
At the same time, the point $P$ is only included in one of four enlarged zones.
So, two shifting subsequences are the same. Following
the enlarging process in an infinite step, we can obtain $\Sigma^1 = \Sigma^2$,
contradicting our original assumption. This contraction
is due to the fact that we have assumed $(\alpha, \beta) \{\Sigma^1\} = (\alpha, \beta) \{\Sigma^2\}$;
thus, $\Sigma^1 \neq \Sigma^2$, then $(\alpha, \beta) \{\Sigma^1\}
\neq (\alpha, \beta) \{\Sigma^2\}$.

For the DNA sequence, some zones in CGR are replenished by points,
so that a pattern appears. In CGR, there exists an correspondence of
more subsequences with different ending $k$-nucleotide strings 
to the same points in bounds of zones. For examples,
subsequences $G^{\infty}A$ in the zone $A$, $T^{\infty}C$ in the zone $C$,
 $A^{\infty}G$ in the zone $G$ and $C^{\infty}T$ in the zone $T$ have the
same points in CGR (1/2,1/2). Under left shift operators, the corresponding relation between
points and subsequences can preserve in zones with small enough lengths.
For example, subsequences $G^{\infty}AC$ in the zone $AC$, $T^{\infty}C^2$ in the zone $C^2$,
 $A^{\infty}GC$ in the zone $GC$ and $C^{\infty}TC$ in the zone $TC$ have the
same points in CGR (1/4,3/4).
In MR of DNA sequences, each zone in CGR is shrunk and clearly divided 
by four bands. There exists a one-to-one correspondence between zones
and ending $k-$nucleotide strings of subsequences.
Frequency of points in the zone can be determined by using MR method as follows. 
 In order to compute frequencies in zones encoded by $k$-nucleotide strings,
 we need to determine partition lines of MR in Fig.~1.
For mononucleotides, there exist $2 \times 2$ zones in the MR. We have $n_1(=3)$ partition lines
$b^1_0=0$, $b^1_1=1/2$ and $b^1_2=1$ along the $\alpha$ axis. For denucleotides, 
there exist $4 \times 4$ zones
in the MR. We have $n_2(=5)$ partition lines $b^2_0=b^1_0=0$, $b^2_1=b^1_1/3=1/6$, $b^2_2=b^1_1=1/2$,
$b^2_3=1-b^2_1=5/6$ and $b^2_4=1-b^2_0=1$ along the $\alpha$ axis. 
In general, for $k-1$-nucleotide strings, if knowing $n_{k-1}(= 2^{k-1} +1)$ partition lines  
$b^{k-1}_i (i=0, 1, \cdots, n_{k-1}-1)$ along the $\alpha$ axis, 
we can obtain $n_k (= 2^k +1=2 n_{k-1}-1)$ partition lines $b^k_i (i=0, 1, \cdots, n_k-1)$  for
$k$-nucleotide strings as follows.
For the $k$-nucleotide strings, there exist $2^k \times 2^k$ zones in the MR. 
The left half $(0 \leq i \leq n_{k-1}-1)$ of partition lines along the $\alpha$ axis 
are described as follows 
\begin{equation}
\begin{array}{ll}
b^k_i = b^{k-1}_{i/2},\,\,\,& {\rm if}\, i\%2 =0; \\
b^k_i = b^{k-1}_i /3,\,\,\, & {\rm if}\, i\%2 =1. \\
\label{eq7}
\end{array}
\end{equation}
From (3), the right half $(n_{k-1} \leq i \leq n_k-1)$ of partition lines along the $\alpha$ axis 
can be determined immediately
\begin{equation}
b^k_i =1- b^k_{n_k-1-i}.
\label{eq8}
\end{equation}
For example, for trinucleotides, 9 partition lines  along the $\alpha$ axis are
$0, \frac{1}{18}, \frac{1}{6}, \frac{5}{18},
\frac{1}{2}, \frac{13}{18},\frac{5}{6},\frac{17}{18}$ and 1.
 We can obtain 17 partition lines  $0,\frac{1}{54},\frac{1}{18}$,
$\frac{5}{54}, \frac{1}{6}, \frac{13}{54}, \frac{5}{18}, \frac{17}{54}$,
$\frac{1}{2},\frac{37}{54},\frac{13}{18},\frac{41}{54},\frac{5}{6}$,
$\frac{49}{54},\frac{17}{18},\frac{53}{54}$ and 1  along the $\alpha$ axis
for tetranucleotides. Partition lines along the $\beta$ axis are the same to 
those along the $\alpha$ axis. Each zone in the MR can thus be surrounded
by the combined partition lines along the $\alpha$ and $\beta$ axes.

Using the MR method, we determine suppression of $k$-nucleotide strings
in HUMHBB (human $\beta$-region, chromosome 11) with 73308 bases 
and YEAST1 (yeast chromosome 1) with 230209 bases in Table I, respectively.
In order to check efficiency of the method, we also determine the number of disappearing
strings in all strings for a giving string length in HUMHBB and YEAST1,
respectively. The results are identical
with those in Table I, respectively. So, the MR method is effective to determine 
suppression of nucleotide strings in DNA sequences.

In CGR of DNA sequences, self-similarity patterns change more obscurely
as lengths of sequences increase. A grey plot describes frequency values
in small zones, which sizes ($2^{-k} \times 2^{-k}$) can be given by
lengths of strings encoding the zones ($k$). Along with increase of
string lengths, the self-similarity patterns in CGR are more clear.
A high and low frequent zones are redivided to smaller and described
by a grey scale. Some empty zones may appear in the patterns of CGR,
i.e., some nucleotide strings are suppressed in the sequences. In the procedure
of decreasing zone sizes, more and more empty zones emerge 
in the patterns of CGR.
For example, evolution of a self-similarity pattern in CGR of
the archaebacteria Archeoglobus fulqidus is shown in Fig.~1 of
Ref. \cite{DGVFF}.
If DNA sequences are infinite, the compositional
structure can be displayed in small enough zones.
Empty zones are a part of the global feature in CGR.
But the DNA sequences are finite. 
A finite sequence, even a random sequence, may also lead to suppression of strings. 
Along with increase of string length, more and more strings
are suppressed in the finite sequences.  

In Table I, we compare the suppression of
nucleotide strings between DNA and random sequences with
the same length.
Suppression of nucleotide strings for HUMHBB starts at $k$=5.
For a random sequence with the same length, which is given by using
a random number generator\cite{PTVF}, suppression of nucleotide strings 
is delayed to start at $k$=7. The number of suppressed nucleotide strings
for the random number is 5.67\% of that for HUMHBB. The finite sequence
 of HUMHBB effects on the suppression of 7-nucleotide strings.
Along with increase of $k$, numbers of suppressed nucleotide strings
for the random number more increase and approach those for HUMHBB.
At $k=$10, the number of suppressed nucleotide strings
for the random number is 99.3\% of that for HUMHBB. In this case, 
suppression of nucleotide strings in HUMHBB is mainly caused 
by the finite length of sequence.
Moreover, suppression of nucleotide strings for YEAST1
starts at $k$=7. For a random sequence with the same length, which is given by using
a random number generator\cite{PTVF}, suppression of nucleotide strings 
is delayed to start at $k$=8. The number of suppressed nucleotide strings
for the random number is 22.7\% of that for YEAST1.
The finite sequence of YEAST1
effects on the suppression of 8-nucleotide strings.
At $k$=10, the  number of suppressed nucleotide strings
for the random number is 97.5\% of that for YEAST1.
Due to the comparison of suppression of nucleotide strings, 
we can thus obtain that HUMHBB and YEAST1 have shorter suppressed nucleotide strings 
than random sequences with the same lengths, respectively. 
Along with increase of string lengths, 
the finite sequences take stronger effects on suppression of nucleotide strings.

In order to display genomic signature, we must eliminate effects of finite sequences
 on suppression of nucleotide strings. 
For a DNA sequence, we take the longest string length before 
suppression of nucleotide strings in a random sequence with the same lengths
as an optimal option of string lengths.
 According to the definition,  string lengths 6 and 7
can be chosen as optimal options for genomic signatures of HUMHBB and YEAST1,
respectively.

\section{Limits of self-similarity scales} 

Suppression of certain nucleotide strings in the DNA sequences leads to
a fractal pattern seen in the MR of DNA sequences.
To quantify the fractal feature in the MR of DNA sequences, 
we introduce information dimension. 
For a given length $k$ of nucleotide strings, we have 
 $M(=N-k+1)$ subsequences $\Sigma_i(i=k,k+1,\cdots, N)$,
 which end with $M$ $k$-nucleotide strings. The subsequences
 are corresponding to $M$ points in a MR.
In the MR,  the length of a zone and the total number of zones
are $\epsilon=3^{-k}$ and $Z=4^k$, respectively.
The numbers of points falling in the $i$-th zone and of non-empty zones
are labeled by $m_i(\epsilon)$ and $Z(\epsilon)$, respectively.
Dividing the  number $m_i(\epsilon)$ by the total point number $M$
 yields a probability $p_i(\epsilon)$ for the $i$-th zone.
Information function and dimension for the points in  MR  
are respectively defined\cite{Farmer} as
\begin{equation}
I(\epsilon)=-\sum_{i=1}^{Z(\epsilon)} p_i {\rm log} p_i,
 \label{eq5}
 \end{equation}
and 
\begin{equation}
D_1= \lim_{\epsilon \rightarrow 0} \frac{I(\epsilon)}{{\rm log} (1/\epsilon)}.
 \label{eq6}
 \end{equation}  
 The information function $I(\epsilon)$ during a range of ${\rm log}(1/\epsilon)$
 has a scaling region. The scaling region reflects the self-similarity of pattern in the MR. 
 The information dimension $D_1$ can be found from the slope 
 in scaling region $I(\epsilon)$ versus ${\rm log}(1/\epsilon)$.
When the length $\epsilon$ of a zone in MR increases from $3^{-k}$ to $2^{-k}$,
 MR of DNA sequences changes to CGR. Information dimension in CGR can thus
 be determined as $({\rm log}_2 3) D_1$. 
We compute information function $I(\epsilon)$ with 
different sizes $\epsilon$ for HUMHBB and draw in Fig.~2.
A linear part of the curve $I(\epsilon)$ versus ${\rm log}(1/\epsilon)$ emerges
between ${\rm log}(1/\epsilon)={\rm log}3=1.10$ and ${\rm log}(1/\epsilon)=6{\rm log}3$=6.59. 
A fitting line is also draw in Fig.~2.
The point for ${\rm log}(1/\epsilon)=7{\rm log}3$=7.69 is started leaving from the line.
Along with the decrease of ${\rm log}(1/\epsilon)$, farther and farther the points
leave from the line.
Since points in the zones correspond to $k$-nucleotide strings, we can obtain that
the self-similarity of pattern in the MR preserves approximately from mononucleotides to
$6$-nucleotide strings, as well as the suppression of many nucleotide strings emerges at 
$7$-nucleotide strings.
Using the least-squares fit method\cite{PTVF} for the liner part, 
we determine its slope, i.e., information dimension $D_1$, to 1.20. 
It is less than the information dimension 1.26 for random sequence with the same length.
Moreover, in Fig.~3, we draw information function $I(\epsilon)$ versus ${\rm log}(1/\epsilon)$ 
for YEAST1.
A linear part of the curve $I(\epsilon)$ versus ${\rm log}(1/\epsilon)$
exists between ${\rm log}(1/\epsilon)={\rm log}3=1.10$ and ${\rm log}(1/\epsilon)=7{\rm log}3$=7.69.
We can obtain that
the suppression of many nucleotide strings in YEAST1 emerges from $8$-nucleotide strings.
Using the least-squares fit method\cite{PTVF} for the liner part, 
we also plot a fitting line in Fig.~3 and determine
 its slope, i.e., information dimension $D_1$, to 1.22.
It is less than the information dimension 1.26 for random sequence with the same length. 
The limits of self-similarity in MR of HUMHBB and YEAST1
are equivalent to the optimal string lengths for genomic signatures, respectively. 
Thus, for presenting genomic signature, a self-similarity limit as
an optimal string length can be determined in computing information dimension.

Using the MR method, we determine suppression of $k$-nucleotide strings
of bacteria complete genomes in Table II, where we put suppression of 
$k$-nucleotide strings in the order of decrease.
For each of the bacteria complete genomes,
a linear part exists in the plot of information function $I(\epsilon)$ versus ${\rm log}(1/\epsilon)$.
From the linear parts, we determine self-similarity limits of genomic signatures in Table II. 
Keeping in the order, we find the suppression of bacteria complete genomes 
does not necessarily depend on the lengths of sequences. The common optimal string length
for the bacteria complete genomic signatures can be chosen as 7.

\section{Conclusion}

In summary, we have shown MR of DNA sequences is one-to-one.
Due to the MR method, suppression of nucleotide strings in the DNA sequences
 is determined. For a DNA sequence,
an optimal string length to display genomic signature 
 is obtained by eliminating effects of the
finite sequence. The optimal string length is further
shown as a self-similarity limit in computing information dimension.
By using the method, self-similarity limits of bacteria complete
genomic signatures are further determined.

\acknowledgments
{ This work was supported in part by the National Key Program for Developing 
Basic Science G1999032801-11.} \\

FIGURES
\\
Fig. 1 Metric representation of HUMHBB. Its boundary and 
partition lines are labeled by solid lines and dash lines, respectively.\\
Fig. 2 A plot of information function $I(\epsilon)$
versus ${\rm log}(1/\epsilon)$ labeled by dots and its fitting line
for HUMHBB.\\
Fig. 3 A plot of information function $I(\epsilon)$
versus ${\rm log}(1/\epsilon)$ labeled by dots and its fitting line
for YEAST1.\\

\begin{table}[htbp]
\label{tab1}
\begin{footnotesize}
Table I. Suppression of $k$-nucleotide strings in HUMHBB, YEAST1 and random sequences.
The total numbers of nucleotide strings for a length $k$ and 
suppressed $k$-nucleotide strings,
 are labeled by $\Pi_k$ and $\Lambda_k$, respectively. 
\begin{tabular}{lcccccc}
$k$                     & 5      &   6   & 7     &   8    &   9     & 10  \\
\tableline
$\Pi_k$                 & 1024     &   4096     & 16384       &  65536       & 262144        & 1048576  \\
\tableline
$\Lambda_k^{HUMHBB}/\Lambda_k^{Random}(73308)$  & 4/0 & 244/0 & 3667/208 & 32909/21402 & 209280/198219 & 985222/977852 \\
\tableline
$\Lambda_k^{YEAST1}/\Lambda_k^{Random}(230209)$ & 0/0 &  0/0  & 110/0  & 8897/2021  & 134302/109290 & 863555/842246 \\
\end{tabular}
\end{footnotesize}
\end{table}

\begin{table}[htbp]
\label{tab2}
\begin{footnotesize}
Table II. Suppression of $k$-nucleotide strings
and self-similarity limits of bacteria complete genomes 
 labeled by $\Lambda_k$ and $k_l$, respectively. 
\begin{tabular}{lcccccc}
$k$          		&   6   & 7     &   8    &   9     & 10  & $k_l$\\
\tableline
$\Lambda_k^{mgen}(580074)$ & 14	    & 851    & 14189 &  126690 &  776767 &7\\
\tableline
$\Lambda_k^{mjan}(1664970)$ &  3    & 318   & 7656  &  84937 &  612138 &8\\
\tableline
$\Lambda_k^{hpyl}(1667867)$  &  2    & 192  & 4290  &  58661 &  538051  &8\\
\tableline
$\Lambda_k^{hpyl99}(1643831)$  &  1    & 130  & 3977   &  58033 &  538512  &8\\
\tableline
$\Lambda_k^{bbur}(910724)$ &  0    & 232   & 8139  &  101444 &  712552 &8\\
\tableline
$\Lambda_k^{rpxx}(1111523)$ &  0    & 71   & 4778    &  79792 &  643520 &8\\
\tableline
$\Lambda_k^{hinf}(1830138)$   &  0    & 12   & 1077    &  33859 &  442423 &8\\
\tableline
$\Lambda_k^{pNGR234}(536165)$   &  0    & 10   & 2881   &  76649 &  699974 &7 \\
\tableline
$\Lambda_k^{mpneu}(816394)$   &  0    & 7   & 2329    &  66513 &  638786  &8\\
\tableline
$\Lambda_k^{mthe}(1751377)$   &  0    & 5   & 665    &  26669 &  408030  &8\\
\tableline
$\Lambda_k^{aquae}(1551335)$   &  0    & 4   & 840    &  33972 &  468735  &8\\
\tableline
$\Lambda_k^{pyro}(1738505)$   &  0    & 4   & 708    &  26863 &  403468  &8\\
\tableline
$\Lambda_k^{aful}(2178400)$   &  0    & 4   & 365    &  16382 &  330488  &8\\
\tableline
$\Lambda_k^{mtub}(4411529)$  &  0    & 3   & 595    &  20793 &  306071  &9\\
\tableline
$\Lambda_k^{pabyssi}(1765118)$   &  0    & 3   & 291    &  18803 &  367742 &8\\
\tableline
$\Lambda_k^{tmar}(1860725)$   &  0    & 2   & 594    &  24329 &  399932  &8\\
\tableline
$\Lambda_k^{cpneu}(1230230)$   &  0    & 2   & 452    &  28569 &  468992  &8\\
\tableline
$\Lambda_k^{ecoli}(4639221)$   &  0    & 1   & 173   &  5595 &  150409  &9\\
\tableline
$\Lambda_k^{synecho}(3573470)$  &  0    & 1   & 149    &  8058 &  214433  &9\\
\tableline
$\Lambda_k^{ctra}(1042519)$   &  0    & 0   & 562    &  34004 &  510293  &8\\
\tableline
$\Lambda_k^{aero}(1669695)$   &  0    & 0   & 137    &  20084 &  401256  &8\\
\tableline
$\Lambda_k^{tpal}(1138011)$   &  0    & 0   & 118    &  20912 &  453066  &8\\
\tableline
$\Lambda_k^{bsub}(4214814)$   &  0    & 0    & 4      &  2919 &  156165   &9\\
\end{tabular}
\end{footnotesize}
\end{table}


\begin{thebibliography}{50}
\bibitem{GM} A. J. Gibbs and G. A. Mcintyre
{\it Eur. I. Biochem.} {\bf 16}, 1 (1970).
\bibitem{IKHOM}  N. Iwabe, K. Kuma, M. Hasegawa, S. Osawa, and T. Miyata
{\it Proc. Natl. Acad. Sci. USA} {\bf 86}, 9355 (1989). 
\bibitem{KMC}  S. Karlin, J. Mrazek, and A. M. Campbell
  {\it J. Bacteriol} {\bf 179}, 3899 (1997).  
\bibitem{Jeffrey} H. J. Jeffrey
  {\it Nucleic Acids Res.} {\bf 18}, 2163 (1990).
\bibitem{HS} K. A. Hill and S. M. Singh
  {\it Genome} {\bf 40}, 342 (1997).
\bibitem{DGVFF}P. J. Deschavanne, A. Giron, J. Vilain, G. Fagot, and B. Fertil 
  {\it Mol. Biol. Evol.} {\bf 16}, 1391 (1999).
\bibitem{Wu}  Z.-B. Wu {\it Electrophoresis} {\bf 21}, 2321 (2000).
\bibitem{Mand} B. B. Mandelbrot {\it The Fractal Geometry of Nature}.
(Freeman and Company, New York, 1983).
\bibitem{PTVF} W. H. Press, S. A. Teukolsky, W. T. Vetterling, and B. P. Flannery
  {\it Numerical Recipes in C}.  2nd ed. (Cambridge University Press, 1992).
\bibitem{Farmer}J. D. Farmer {\it Physica D} {\bf 4}, 366 (1982).
\end{thebibliography}
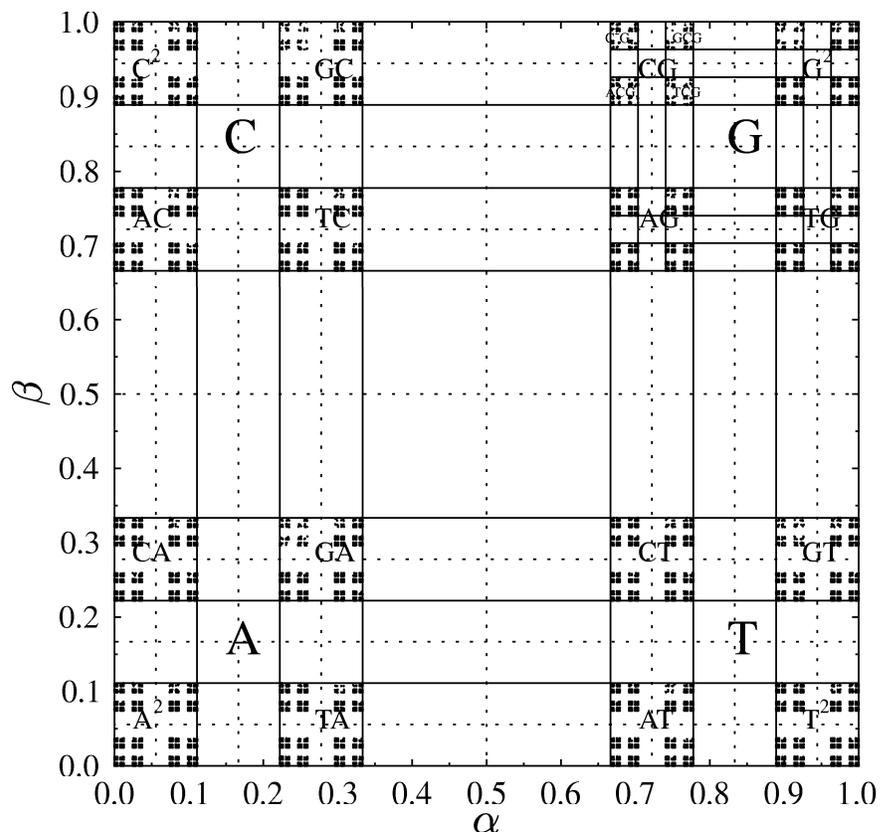
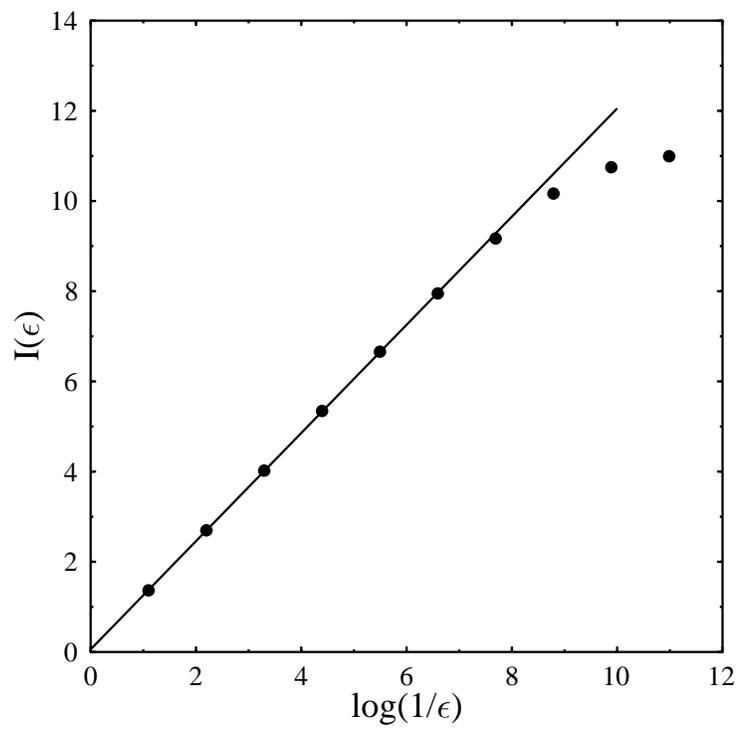
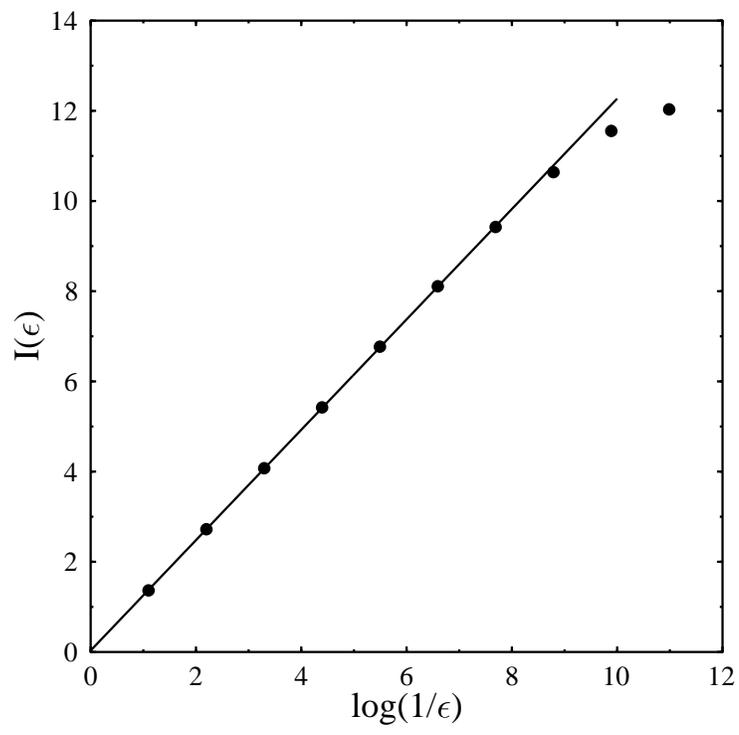
\end{document}